\documentclass[aps,prl,reprint,floatfix,citeautoscript,longbibliography,superscriptaddress]{revtex4-1}
\usepackage{graphicx}
\usepackage{bm,amsmath,amssymb,mathrsfs,dcolumn}
\usepackage{gensymb}
\usepackage[usenames]{color}
\usepackage{subfigure}
\usepackage{ulem}
\usepackage{soul,xcolor}

\newcommand{\Heff}{\ensuremath{\mathbf{H}_\mathrm{eff}}}

\begin{document}

\title{Classification of Magnetic Forces on Antiferromagnetic Domain Wall}
%\title{Manipulation of antiferromagnetic domain wall by magnetic field}
%\title{Route to fast moving antiferromagnetic domain wall}
\author{H. Y. Yuan}
\altaffiliation{These authors contributed equally to this work.}
\affiliation{Department of Physics, Southern University of Science
and Technology, Shenzhen 518055, Guangdong, China}
\author{Weiwei Wang}
\altaffiliation{These authors contributed equally to this work.}
\affiliation{Department of Physics, Ningbo University, Ningbo 315211, China}
\author{Man-Hong Yung}
\email[Electronic address: ]{yung@sustc.edu.cn}
\affiliation{Institute for Quantum Science and Engineering and Department
of Physics, Southern University of Science and Technology, Shenzhen, 518055, China}
\affiliation{Shenzhen Key Laboratory of Quantum Science and Engineering, Shenzhen, 518055, China}
\author{X. R. Wang}
\email[Electronic address: ]{phxwan@ust.hk}
\affiliation{Department of Physics, The Hong Kong University of
Science and Technology, Clear Water Bay, Kowloon, Hong Kong}
\affiliation{HKUST Shenzhen Research Institute, Shenzhen 518057, China}
\date{\today}

\setstcolor{red}

\begin{abstract}
A major challenge in spintronics is to find an efficient means to
manipulate antiferromagnet (AFM) states, which are inert relative to
a uniform magnetic field, due to the vanishingly-small net magnetization.
The question is, how does an AFM response to an inhomogeneous field?
Here we address the problem through a complete classification of the
magnetic forces on an AFM domain wall (DW), revealing the following
physical properties:
(i) the tiny net magnetization still responses to the field gradient.
(ii) the N\'{e}el order is sensitive to the field difference between
two sublattices.
(iii) DW energy has a quadratic dependence on the magnetic field due
to its noncollinear structure. Remarkably, the first two factors
drive DW to the opposite directions in a nanowire, but the
third effect tends to push the DW to the high field region. Consequently,
the competition among these three forces  can be applied to understand the
seemingly-contradictory results on AFM motion in literature.
Additionally, our results provide a new route for a speedy manipulating AFM DW;
our numerical simulation indicated that for a synthetic antiferromagnet, the
DW propagating speed can reach tens of kilometers per second, an order of
magnitude higher than that driven by an electric current.
\end{abstract}

\maketitle
%\section{Introduction}
{\it Introduction.---} Antiferromagnets (AFMs), being promising for
spintronics, have attracted much attention for research in recent years
~\cite{Kimel2004,Duine2007,Haney2008,Xu2008, Kampfrath2010, Mac2011,Hals2011,
Tveten2013,Cheng2014, Marti2014, Wadley2016,Helen2016,Shiino2016,Selzer2016,
Jungwirth2016, Xichao2016, Barker2016, Fukami2016, Yuan2017, Wei2017},
due to the superior stability and terahertz spin dynamics.
However, at the same time, the high stability of AFM also represents a
problem for controlling AFM states.
%AFMs, due to the zero/vanishingly small net magnetization, are
%insensitive to many external forces and interact hardly with surrounding
%systems with one known exception of exchange bias at AFM/ferromagnet
%interfaces.
Various methods, including electric currents through spin-transfer
torque~\cite{Hals2011}, spin-orbit torque \cite{Helen2016,
Shiino2016}, spin waves~\cite{Tveten2014}, and thermal gradients
\cite{Selzer2016,Takei2014,Yuan2017bec,Wu2016} have been proposed to manipulate
AFM states, and in particular domain wall (DW) dynamics.
However, each of these proposals has its own limitations.
For example, although the spin-orbit torque can drive a DW propagating
at a speed of up to several kilometers per second and is free of Walker
breakdown~\cite{Helen2016}, its application is limited to metals with a
broken inversion symmetry, which excludes a large number of normal AFMs.
On the other hand, for thermally-driven DW motion, the underlying physical
mechanism remains unclear. Overall, finding efficient ways to manipulate
AFM states remains a fundamental problem and is of crucial importance
for a variety of applications.

Remarkably, determination of the DW propagating direction has become
a debate topic \cite{Schlickeiser2014,Xiansi2014,Peng2015}. One recent
observation~\cite{Tveten2016} is that AFM DWs do have a non-zero
magnetization; they can interact with a magnetic field through the
Zeeman effect. Consequently, it was first predicted that the DW
velocity should be anti-symmetric relative to the applied magnetic
field. However, this prediction was later challenged by another
theory~\cite{Helen2016}, which proposed a quadratic Zeeman energy
for an AFM DW; in this way, the field-dependence of DW velocity
should be symmetric in the field direction instead. So, which one is
correct? In fact, our numerical atomistic simulations indicate that
neither of these two predictions are consistent in general, which
calls for a need for a further investigation.
%suggesting the necessity to recover the full physics in magnetic field driven AFM DW motion.

In this paper, we thoughtfully study the AFM DW motion under inhomogeneous
fields, where we classify three DW driving forces based on the following
observations. (i) The net magnetization of an AFM DW interacts with the field
gradient of the nearest unit-cells as reported in Ref.~\cite{Tveten2016}.
(ii) In each unit cell, the N\'{e}el order interacts with the
magnetic-field difference between the two sublattices, where the
spin-orbit field~\cite{Helen2016} can be regarded as a special case.
Note that the these two forces (i) and (ii) have a linear dependence
in, respectively, the field gradient among unit-cells and field
difference within each unit-cell. Furthermore, they tend to cancel
each other. Finally, (iii) AFM DW energy depends on the average
magnetic field quadratically due to the non-collinear structures.

 Overall, the moving direction of an AFM DW depends on the interplay
 of all three mechanisms. These results provide an explanation on
 why our numerical results indicating that DW velocity is neither
 symmetric nor asymmetric, resolving the apparent inconsistency
 between the results of Ref.~\cite{Helen2016} and~\cite{Tveten2016}.
 In addition, our classification of the forces points to a solution
 to designing a spatial profile on the magnetic field that can
 potentially drive an AFM DW to an unprecedentedly-high speed.

\begin{figure}
\centering
\includegraphics[width=0.45\textwidth]{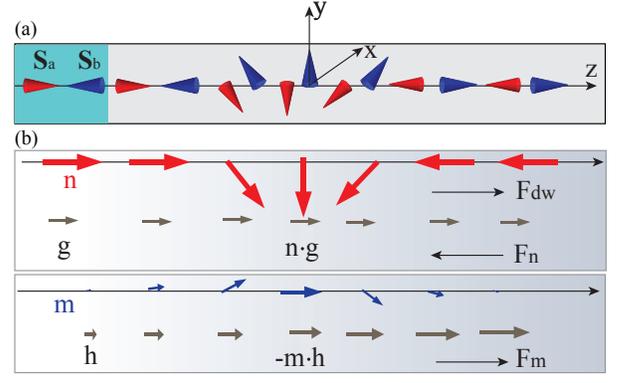}\\
\caption{(Color online)(a) Schematics of a two-sublattice AFM
nanowire with a head-to-head $180^\circ$ domain wall.
(b) Schematics of a spatial distribution of N\'{e}el order $\mathbf{n}$
(red arrow) and magnetization order $\mathbf{m}$
(blue arrow) of a $180^\circ$ DW. The directions of the three
driven forces are also indicated as thin arrows.}
\label{fig1}
\end{figure}

{\it Theory.---} Let us start with a two-sublattice AFM uniaxial nanowire
along the $z$ axis as shown in Fig. \ref{fig1}(a). The system
is described by the following Hamiltonian,
\begin{equation}
\begin{aligned}
\mathcal{H}=& J \sum_{\langle i,j \rangle} \mathbf{S}_{ai}\cdot
\mathbf{S}_{bj}-K_z \sum_i (\mathbf{S}_{ai,z}^2 +\mathbf{S}_{bi,z}^2) \\
&-\sum_i(\mathbf{S}_{ai}\cdot\mathbf{H}_a+\mathbf{S}_{bi} \cdot
\mathbf{H}_b),
\label{hein}
\end{aligned}
\end{equation}
where $\mathbf{S}_{ai}$ ($\mathbf{S}_{bj}$) ($|\mathbf{S}_{ai}|
=|\mathbf{S}_{bj}|=S$) are the spins on sublattices $a$ ($b$).
$\langle i,j \rangle$ denotes the nearest-neighboring sites.
$\mathbf{H}_a$ and $\mathbf{H}_b$ are respectively the external
field on sublattices $a$ and $b$. The first and second terms in
Eq. (\ref{hein}) are the exchange energy (coefficient $J>0$)
and anisotropy energy (coefficient $K_z >0$), respectively.
The third term is the usual Zeeman energy.

In terms of the magnetization, $\mathbf{m}\equiv (\mathbf{S}_{ai}+
\mathbf{S}_{bi})/2S$, and N\'{e}el order, $\mathbf{n}\equiv (\mathbf{S}_{ai}- \mathbf{S}_{bi})/2S$,
the Hamiltonian density $\mathcal{H}$ in the continuum limit
is given by,~\cite{Tveten2013,Tveten2016}
\begin{equation}
\mathcal{H}= \frac{a}{2} \mathbf{m}^2 + \frac{A}{2} (\partial_z\mathbf{n})^2
-\frac{K}{2} n_z^2+ L  \mathbf{m} \cdot \partial_z \mathbf{n}
- \mathbf{m} \cdot \mathbf{h} + \mathbf{n} \cdot \mathbf{g}\ ,
 \label{fe}
\end{equation}
where $a\equiv 4JS^2$ and $A\equiv d^2 JS^2$ are, respectively,
homogeneous and inhomogeneous exchange constants with $d$ being the
lattice constant. $K=4K_zS^2$.
$\mathbf{h}\equiv (\mathbf{H}_a+\mathbf{H}_b)/2$ and
$\mathbf{g}\equiv (\mathbf{H}_a-\mathbf{H}_b)/2$ are, respectively,
the average field (over a unit cell) and field difference on the two
sublattices (in a unit-cell). The term containing
$L \equiv 2dJS^2$ breaks the parity symmetry and results in a net
magnetic moment of an AFM DW~\cite{Tveten2016}.

Consequently, the dynamics of the order parameters, $\mathbf{m}$
and $\mathbf{n}$, are governed by the following equation of
motion~\cite{Hals2011,Yuan20171},
\begin{align}
&\partial_t \mathbf n =-\mathbf n  \times \left( \mathbf h_m-
\alpha_m \partial_t \mathbf m \right),\nonumber\\
&\partial_t \mathbf m =-\mathbf n \times \left(\mathbf h_n-
\alpha_n\partial_t\mathbf n   \right)-\mathbf m \times
\left(\mathbf h_m-\alpha_m \partial_t\mathbf m \right ),
\label{mneq}
\end{align}
where $\mathbf{h}_m \equiv-\delta \mathcal{H}/\delta\mathbf{m}$
and $\mathbf{h}_n\equiv -\delta\mathcal{H}/ \delta \mathbf{n}$
are, respectively, the effective fields on the magnetization order
$\mathbf{m}$ and N\'{e}el order $\mathbf{n}$.
One of the main contribution to $\mathbf{h}_n$ is the field
difference on the two sublattices. $\alpha_m$ and $\alpha_n$ are
the damping coefficients associated with the motion of $\mathbf{m}$
and $\mathbf{n}$, which can be determined from the first-principles
theory~\cite{Yuan20171}.

To the first order in $\mathbf{m}$, one can decouple the motion
from Eq.~(\ref{mneq}) and obtain,
\begin{equation}
\mathbf{n} \times[ \partial_{tt} \mathbf{n} +a \alpha\partial_t
\mathbf{n}+ \bm{T}_s+\bm{T}_d ] =0 \ ,
\label{neq}
\end{equation}
where $\bm{T}_s\equiv-a(A-L^2/a)\partial_{zz}\mathbf n-(aK-h^2)
n_z \hat{\bm{z}}$, $\bm{T}_d\equiv 2L\partial_z\mathbf n\times
\partial_t \mathbf n+\mathbf h \times\partial_t\mathbf n  - L
\partial_z \mathbf h+ a \mathbf{g}$, and $\alpha\equiv\alpha_n$.
Note that for the motion of wide DWs, the $\alpha_m$ term can
be neglected~\cite{Yuan20172}, even though its value can be
significantly larger than $\alpha_n$.

{\it AFM domain wall.---} First of all, the static DW profile can be obtained
by setting $\partial_t \mathbf n = 0$ in Eq.~(\ref{neq}).
In spherical coordinates, we denote $\mathbf{n}=(\sin\theta \cos\phi,
\sin\theta\sin\phi, \cos \theta)$. Similar to the ferromagnetic counterpart~\cite{Yuan2016},
we apply the following DW ansatz: $\theta =2\tan^{-1}
\{\exp[(z-z_c)/\Delta]\}$, where $z_c$ is the DW center and $\Delta$
the DW width. For a uniform field ($\mathbf{g}=0$), the solution is,
\begin{equation}\label{DMWwidth}
\Delta ={\Delta _0}/{\left( {1-{h^2}/{aK}} \right)^{1/2}} \ ,
\end{equation}
where $\Delta_0 \equiv \sqrt{A/2K}$ is the DW width in the absence
of the external field. Therefore, the DW width increases with the
applied field, but breaks down at the critical field ${h_c} = \sqrt {aK}$.

Next, for a propagating DW, we apply the ansatz of $\mathbf{n}
=\mathbf{n}(z-vt)$, where $v=\partial_t z_c$ is the velocity of DWs.
From the vector product, $\partial_z \mathbf{n} \times \mathrm{Eq}.
\ (\ref{neq})$, followed by an integration over the whole space, one can obtain
the Thiele equation \cite{Thiele1972, Tveten2014},
\begin{equation}
M_{zz} ( \partial_{tt} z_c + a \alpha \partial_t z_c ) = F_z,
\end{equation}
where $M_{zz}\equiv(1/a)\int (\partial_z\mathbf{n})^2dz=2/(a \Delta)$.
Here $F_z$ is the effective driving force on the DW. We found that
for any applied field, the driving force can always be decomposed
into three different components, i.e., $F_z=F_m+F_n+F_{dw}$, where,
\begin{equation}
\begin{aligned}
&F_m =\frac{ L}{a\Delta}\int({\partial_z}h)\sin^2\theta\ dz \ ,\\
&F_n=-\frac{1}{\Delta}\int g \sin^2\theta \ dz \ ,\\
&F_{dw}=\frac{1}{2A}\int ( \partial_z h^2)\sin^2\theta dz \ .\\
\end{aligned}
\label{threeF}
\end{equation}
For a steady DW motion where $\partial_{tt} z_c =0$, the velocity
of the DW depends on the relative sizes of these forces,
\begin{equation}
v = (\Delta /2\alpha )\left(F_m+F_n+ F_{dw} \right) .
\label{th_v}
\end{equation}
As a result, the propagation direction of the DW depends on the
relative sizes of the force components. In other words, the velocity
can be symmetric, anti-symmetric and asymmetric, depending on the
spatial profile of the applied field; this framework can be applied
to understand the conflicting results between Ref~\cite{Helen2016}
and \cite{Tveten2016}.

To elaborate further, let us now discuss the physical origins of
the these three forces. First, $F_m$ comes from the net magnetization
of a DW originated from the parity-breaking term ($L$).
This net magnetization interacts with the external field,
and can sense the average magnetic field gradient.
As a result, the DW tends to move to the high field region to reduce
the total Zeeman energy $-\mathbf{m}\cdot\mathbf{h}$, i.e. to the
direction of $\partial h/\partial z>0$, as shown in the bottom panel
of Fig.~\ref{fig1}(b). Second, $F_n$ is proportional to the average
field difference on the two sublattices, arising from the spatial
variation of the magnetic field. It plays the role of a N\'{e}el
field \cite{Zelezny2014} that couples to the N\'{e}el order as
indicated by the term of $\mathbf{n \cdot g}$ in Eq \eqref{fe}.
For a continuous monotonic spatially varying field,
$F_n$ is related to the field derivative and has an opposite sign to $F_m$.
This force tends to drive a DW to move along the direction of
$\partial h/\partial z<0$. Finally, the force $F_{dw}$ is referred to
as the ``field-dependent DW energy effect", based on the fact that
$K$ is modulated by a factor $1-h^2/aK$ as shown in the DW width
given by Eq.~(\ref{DMWwidth}).
A DW tends to move to the direction with smaller anisotropy i.e.
larger $h^2$ region, to reduce the total free energy.
Therefore the reversal of the field direction does not change the
direction of DW motion.

\begin{figure}
\centering
\includegraphics[width=0.4\textwidth]{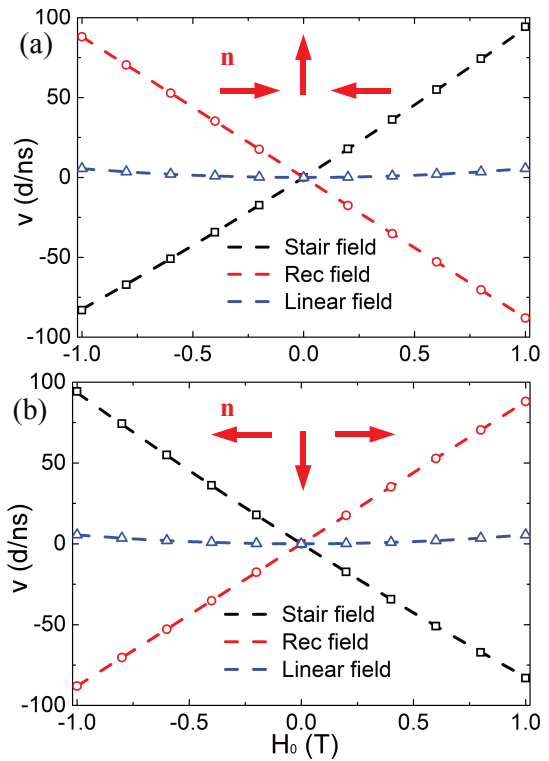}\\
\caption{(Color online)
Velocity of a head-to-head DW (a) and a tail-to-tail DW (b) as a
function of field stength for the linear (triangles), stair (squares)
and rectangle (circles) fields, respectively. The dashed lines are
theoretical prediction Eq.~(\ref{th_v}). The model parameters are
$N=10^3$, $J=16$ meV, $K=0.02$ meV, $S=1$ and $\alpha = 0.02$.}
\label{fig2}
\end{figure}

{\it Examples.---} Next, we shall show that the effects of the three
forces can be manifested independently, by considering three different
types of inhomogeneous external fields, namely,
\begin{align}
&\mathsf{Linear:}\ H_i = H_0 \frac{i}{2N},(i=0,1,...2N-1) \nonumber \\
&\mathsf{Stair:}\ H_{2i}=H_{2i+1}=H_0 \frac{i}{N},(i=0,1,...N-1)  \\
&\mathsf{Rectangular:}\ H_{2i}=0, H_{2i+1}=\frac{H_0}{2N}, (i=0,1,...N-1)\nonumber
\end{align}
where $2N$ is the total number of spins and $H_0$ characterizes the
field inhomogeneity. For the linear field, the $F_{dw}$ term dominates the DW
motion, which means that the velocity does not change if the applied
field is reversed. It is because $L/A=d/2,g = d\partial_z h/2$, the first
two forces cancel each other, i.e., $F_m = -F_n$. In this case, the
velocity in Eq.~(\ref{th_v}) can be approximated~\cite{note02}
by $v \approx  \Delta^2H_0^2/(2a\alpha N)$, which is shown with the
blue line in Fig.~\ref{fig2}(a). The value is normally very small
(order of tens meter per second at most) for a typical field gradient
of $1~\mathrm{T/\mu m}$. For the stair field, $F_n=0$ and $F_m$
dominates the DW motion. In the weak field regime $H_0 \ll L/\Delta$,
$F_{dw}$ gives a second order correction to the velocity and thus
we have $v= L \Delta H_0/(a \alpha N)+\Delta^2H_0^2/(a\alpha N)$,
which is shown by the black line in Fig.~\ref{fig2}(a).
Here the moving direction of DW is reversed when the field is reversed.
Note that the magnitude of the velocity is asymmetric to the external
field due to the contribution from $F_{dw}$.

For the rectangular field, $F_n$ dominates DW motion because $F_m= 0$.
The DW velocity is $v=\Delta H_0/(\alpha N)$, as shown with
the red line in Fig.~\ref{fig2}(a). In this case, the moving direction is
reversed as the field reverses. Moreover, a tail-to-tail DW moves in
the opposite direction to that of a head-to-head DW under stair and
rectangle fields while the moving direction does not depend on
DW types under linear fields, as shown in Fig. \ref{fig2}(b).
This feature can be readily explained by analyzing the direction
of the three forces.

To verify our analytical results for the three scenarios, we performed
numerical simulations of coupled Landau-Lifshitz-Gilbert (LLG)
equations for the two sublattice spins,
$\partial_t \mathbf{S}_i = - \mathbf{S}_i \times \Heff + (\alpha/S )
\mathbf{S}_i \times  \partial_t \mathbf{S}_i$, where $\alpha$ is the
Gilbert damping and $\Heff$ is the effective field given by
$\Heff = - ({\delta \mathcal{H}}/{\delta \mathbf{S}_i})$.
In Fig.~\ref{fig2}, the black rectangles, red circles and blue triangles
represent the field dependence of DW velocity for linear, stair, and
rectangular fields, respectively. It is shown that the theoretical
predictions [Eq.~(\ref{th_v})] agree very well with the numerical simulations.

{\it High-speed DW on synthetic AFMs---} Within this framework of
classifying the forces on AFMs, we are now ready to discuss how an AFM
DW can be manipulated efficiently. According to above analysis, it
is clear that the quadratic force $F_{dw}$ is only a second-order effect
for a weak field, and $F_m$ is proportional to the gradient of average
magnetic field that is normally not very large. Therefore, in order
to achieve a high DW velocity, one can instead consider strengthening
the the inter-sublattice force $F_n$, i.e., with a large average-field
difference on the two sublattices.

\begin{figure}
\centering
\includegraphics[width=0.4\textwidth]{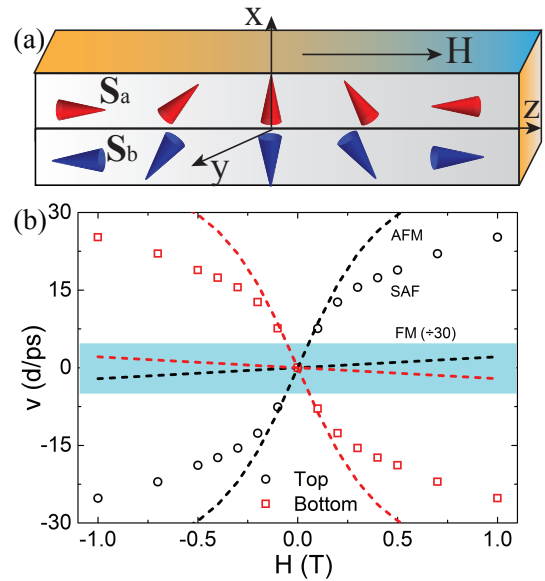}\\
\caption{(Color online) (a) Schematic illustration of a synthetic
antiferromagnet consists of two ferromagnetic chains with an
antiferromagnetic inter-chain coupling. (b) Simulation results of
DW velocity as a function of the applied field. The dashed line
labelled with AFM is for a traditional two-sublattice a synthetic
AFM under a rectangle field while the dashed line labelled with
FM is for a single ferromagnetic chain. The $\div 30$ sign means
that the real DW velocity is 1/30 of the plot. The intra-layler
and inter-layer exchange coupling are $J_1 = -16$ meV and
$J_2 = 16$ meV, respectively. $K=0.02$ meV, $\alpha=0.02$.}
\label{fig3}
\end{figure}

Specifically, we predict that a strong inter-sublattice force can be
achieved readily in a synthetic antiferromagnets (SAFM) \cite{Parkin1990,Bennet1990},
which consists two antiferromagnetically coupled ferromagnetic chains
(see Fig.~\ref{fig3}(a)). When the inter-chain coupling is much stronger
than the intra-chain exchange coupling, this system is equivalent to
a one-dimensional antiferromagnets, where the atoms on the top and
bottom chains correspond to the sublattices $a$ and $b$, respectively.
If a uniform field is applied in the top/bottom chain, its effect
corresponds to a rectangle field applied in the two sublattices,
and this could effectively drive the coupled DW to move.

As shown in Fig.~\ref{fig3}(b), our numerical simulations show that
DW velocity increases with the field, and reaches a value of 17d/ps
(around 10 km/s) for a 0.5 T field. This is almost one order of
magnitude larger than the velocity of electric-current-driven DW
motion (0.75 km/s)~\cite{Yang2015}. For a comparison, the DW velocity
in the system without inter-layer AFM coupling is shown as the dashed
lines and it is almost three orders of magnitude smaller than the
velocity in a SAFM. This is because the velocity is given by,
$v_\mathrm{FM}= H\Delta(\alpha+1/\alpha)\approx H \Delta\alpha$ \cite{Wieser2010},
while the velocity in an SAFM is scaled by a factor of $1/\alpha$.
Given that $\alpha \ll 1$, the DW can move much faster in a SAFM.

Therefore, the synthetic antiferromagnets is a promising system for the high
DW velocity of around 10 km/s in a uniform field in one of the
ferromagnetic layers. This mechanism should work for metals,
insulators, and semiconductors. The driving field can be the
Oersted field generated from an electric current. Alternatively,
the spin-orbit field in the two antiferromagnetically coupled
ferromagnetic layer sandwiched by a non-magnetic layer also takes
opposite signs at the two interfaces and
thus can be used to induce fast DW motion.

{\it Discussions and Conclusions.---} Finally, let us compare our
results with the conflicting theoretical predictions in the literature.
If the net magnetization dominates the DW motion, then the DW propagation
indeed reverses its direction as the field direction is reversed.
The simulation results in literature~\cite{Tveten2016}, based on the
dynamic equations of $\mathbf{m}$ and $\mathbf{n}$, did not include the
 N\'{e}el field term of $\mathbf{n}\cdot \mathbf{g}$. In fact, their
``linear field" corresponds to our stair field; the DW propagating
direction should reverse as the field direction is reversed. However,
if the linear field is considered in a Heisenberg model, the DW velocity
should be symmetric with respect to the field, since $F_m$
and $F_n$ cancel each other and the dominate driven force is DW energy,
which depends quadratically on magnetic fields. These observations
are well justified by our atomistic numerical simulations.

Furthermore, we note that the quadratic field-dependence of DW energy
is fundamentally different from previous quadratic-field term from the
interaction between the external field with its induced magnetization~\cite{Helen2016}.
In an AFM, the magnetic susceptibility is zero or vanishingly small
when the temperature is far below the N\'{e}el temperature and the field
is smaller than the spin-flop field \cite{Trapp1962, Coey2009}.
Thus the effect caused by the induced magnetization should not
play an important role in DW motion.

In conclusion, we provided a framework for studying AFM DW motion under
the spatial inhomogeneous magnetic fields, where any applied field can
be decomposed into three different force components, for driving an AFM
DW. The three force components are respectively originated from the
net DW magnetization that interacts with an averaged field over the
neighboring unit-cells, the field difference on the two sublattices
that couples to the N\'{e}el order and plays a role of the N\'{e}el
field, and quadratic field dependence of DW energy due to the non-collinear
DW spin structure. The first two forces tend to cancel each other for
a linear field. The third force can drive a DW move at a speed that
is insensitive to both DW types and field direction. To produce a
high-speed DW motion, rectangular or stair field is favorable for
taking the advantage of the second force. Finally, supported by
numerical simulations, we predicted that SAFM can become a promising
candidate for realizing this proposal.

\begin{acknowledgments}
{\it Acknowledgments.---}
We acknowledge the financial support from National Natural Science
Foundation of China (NSFC) Grants (Nos. 61704071 and 11604169).
MHY acknowledges support by NSFC Grant (No. 11405093), Guangdong
Innovative and Entrepreneurial Research Team Program (2016ZT06D348),
and Science, Technology and Innovation Commission of Shenzhen
Municipality (ZDSYS20170303165926217 and JCYJ20170412152620376).
XRW was supported by the NSFC Grant (No. 11774296) as well
as Hong Kong RGC Grants (Nos. 16300117 and 16301816).
\end{acknowledgments}


\begin{thebibliography}{}
\bibitem{Kimel2004} A. V. Kimel, A. Kirilyuk, A. Tsvetkov, R. V. Pisarev,
and Th. Rasing, Nature(London) \textbf{429}, 850 (2004).

\bibitem{Duine2007} R. A. Duine, P. M. Haney, A. S. Nunez, and
A. H. MacDonald, Phys. Rev. B {\bf 75}, 014433 (2007).

\bibitem{Haney2008} P. M. Haney and A. H. MacDonald,
Phys. Rev. Lett. \textbf{100}, 196801 (2008).

\bibitem{Xu2008} Y. Xu, S. Wang, and K. Xia, Phys. Rev. Lett.
\textbf{100}, 226602 (2008).

\bibitem{Kampfrath2010} T. Kampfrath, A. Sell, G. Klatt, A. Pashkin,
S. M\"{a}hrlein, T. Dekorsy, M. Wolf, M. Fiebig, A. Leitenstorfer,
and R. Huber, Nat. Photon. \textbf{5}, 31 (2010).

\bibitem{Mac2011} A. H. MacDonald and M. Tsoi, Philos. Trans. R. Soc. A
\textbf{269}, 3098 (2011).

\bibitem{Hals2011} K. M. D. Hals, Y. Tserkovnyak, and A. Brataas,
Phys. Rev. Lett. \textbf{106}, 107206 (2011).

\bibitem{Tveten2013} E. G. Tveten, A. Qaiumzadeh, O. A. Tretiakov,
and A. Brataas, Phys. Rev. Lett. \textbf{110}, 127208 (2013).

\bibitem{Cheng2014} R. Cheng, J. Xiao, Q. Niu, and A. Brataas,
Phys. Rev. Lett. \textbf{113}, 057601 (2014).

\bibitem{Marti2014} X. Marti et al., Nat. Mater. \textbf{13}, 367 (2014).

\bibitem{Wadley2016} P. Wadley et al., Science \textbf{351}, 587 (2016).

\bibitem{Helen2016} O. Gomonay, T. Jungwirth, and J. Sinova,
Phys. Rev. Lett. \textbf{117}, 017202 (2016).

\bibitem{Shiino2016} T. Shiino, S. H. Oh, P. M. Haney, S. -W. Lee, G. Go,
B. -G. Park, and K. -J. Lee, Phys. Rev. Lett. \textbf{117}, 087203 (2016).

\bibitem{Selzer2016} S. Selzer, U. Atxitia, U. Ritzmann, D. Hinzke,
and U. Nowak, Phys. Rev. Lett. \textbf{117}, 107201 (2016).

\bibitem{Jungwirth2016} T. Jungwirth, X. Marti, P. Wadley, and J. Wunderlich,
Nat. Nanotech. \textbf{11}, 231 (2016).

\bibitem{Xichao2016} X. Zhang, Y. Zhou, and M. Ezawa,
Sci. Rep. \textbf{6}, 24795 (2016).

\bibitem{Barker2016} J. Barker and O. A. Tretiakov,
Phys. Rev. Lett. \textbf{116}, 147203 (2016).

\bibitem{Fukami2016} S. Fukami, C. Zhang, S. Dutta Gupta, A. Kurenkov,
and H. Ohno, Nat. Mater. \textbf{15}, 535 (2016).

\bibitem{Yuan2017} H. Y. Yuan and X. R. Wang, Appl. Phys. Lett.
\textbf{110}, 082403 (2017).

\bibitem{Wei2017}W. Wang, C. Gu, Y. Zhou, and H. Fangohr,
Phys. Rev. B \textbf{96}, 024430 (2017).
%\cite{Kimel2004,Duine2007,Haney2008,Xu2008, Kampfrath2010, Mac2011,Hals2011,Tveten2013,
%Cheng2014, Marti2014, Wadley2016,Helen2016,Shiino2016,Selzer2016,Jungwirth2016, Barker2016, Fukami2016, Yuan2017}

\bibitem{Tveten2014} E. G. Tveten, A. Oaiumzadeh, and A. Brataas,
Phys. Rev. Lett. \textbf{112}, 147204 (2014).

\bibitem{Takei2014} S. Takei, B. I. Halperin, A. Yacoby, and Y. Tserkovnyak,
Phys. Rev. B \textbf{90}, 094408 (2014).

\bibitem{Yuan2017bec} H. Y. Yuan and M.-H. Yung, preprint at arXiv:1711.04394

\bibitem{Wu2016} S. M. Wu et al.,  Phys. Rev. Lett. \textbf{116}, 097204 (2016).

\bibitem{Schlickeiser2014} F. Schlickeiser, U. Ritzmann, D. Hinzke,
and U. Nowak, Phys. Rev. Lett. \textbf{113}, 097201 (2014).

\bibitem{Xiansi2014} X. S. Wang and X. R. Wang, Phys. Rev. B
\textbf{90}, 014414 (2014).

\bibitem{Peng2015}P. Yan, Y. Cao, and J. Sinova,
\textbf{92}, 100408(R) (2015).

%\bibitem{Keffer1952} F. Keffer and C. Kittel,
%Phys. Rev. \textbf{85}, 329 (1952).

\bibitem{Tveten2016} E. G. Tveten, T. Muller, J. Linder, and A. Brataas,
Phys. Rev. B \textbf{93}, 104408 (2016).

\bibitem{Yuan20171} Q. Liu, H. Y. Yuan, K. Xia, and Z. Yuan,
Phys. Rev. Mater. \textbf{1}, 061401(R) (2017).

\bibitem{Yuan20172} H. Y. Yuan, Q. Liu, K. Xia, Z. Yuan and X. R. Wang (unpublished).

\bibitem{Yuan2016} H. Y. Yuan, Z. Yuan, K. Xia, and X. R. Wang,
Phys. Rev. B \textbf{94}, 064415 (2016).

\bibitem{Thiele1972} A. A. Thiele, Phys. Rev. Lett. \textbf{30}, 230 (1972).

\bibitem{Zelezny2014} J. \v{Z}elezn\'{y}, H. Gao, K. V\'{y}born\'{y}, J. Zemen,
J. Ma\v{s}ek, A. Manchon, J. Wunderlich, J. Sinova, and T. Jungwirth,
Phys. Rev. Lett. \textbf{113}, 157201 (2014).

\bibitem{note02} The domain wall velocity is averaged on the first
1 ns. Since the domain wall speed is very small in this period, we assume
that the domain wall center is close to its initial position such that the
integral in $F_{\mathrm{dw}}$ can be calculated analytically to derive
the DW velocity.

%\bibitem{note03}If one could produce a stair field locating at the
%region of domain walls (several nm to tnes of nm), the maxium velocity
%driven by inte-sublattice force can be increased.

\bibitem{Parkin1990} S. S. P. Parkin, N. More, and K. P. Roche,
Phys. Rev. Lett. \textbf{64}, 2304 (1990).

\bibitem{Bennet1990} W. R. Bennett, W. Schwarzacher, and W. F. Egelhoff,
Phys. Rev. Lett. \textbf{65}, 3169 (1990).

\bibitem{Yang2015} S. -H. Yang, K. -S. Ryu, and S. Parkin,
Nat. Nanotech. \textbf{10}, 221 (2015).

\bibitem{Trapp1962} C. Trapp and J. W. Stout, Phys. Rev. Lett. \textbf{10}, 157 (1962).

\bibitem{Coey2009} J. M. D. Coey, Magnetism and Magnetic Materials,
Cambridge Unviersity Press (New York, 2010)

\bibitem{Wieser2010} R. Wieser, E. Y. Vedmedenko, and R. Wiesendanger,
Phys. Rev. B \textbf{81}, 024405 (2010).
\end{thebibliography}
\end{document}